\documentclass[letterpaper,preprint,aps]{revtex4-1}
\usepackage[latin9]{inputenc}
\usepackage{verbatim}
\usepackage{amsmath}
\usepackage{amssymb}
\usepackage[unicode=true,pdfusetitle,
 bookmarks=true,bookmarksnumbered=false,bookmarksopen=false,
 breaklinks=false,pdfborder={0 0 1},backref=false,colorlinks=false]
 {hyperref}

\makeatletter

\pdfpageheight\paperheight
\pdfpagewidth\paperwidth

\usepackage{graphicx}
\usepackage{tikz}
\usepackage{bbm}
\usepackage{verbatim}
\usepackage[percent]{overpic}

\DeclareMathOperator{\sech}{sech}

\def\dd{{\rm d}}

\makeatother

\begin{document}
\title{Conformal Wave Expansions for Flat Space Amplitudes}
\author{Chang Liu}
\email{chang\_liu3@brown.edu}

\author{David A. Lowe}
\email{lowe@brown.edu}

\affiliation{Department of Physics, Brown University, Providence, RI, 02912, USA}
\begin{abstract}
The extended BMS algebra contains a conformal subgroup that acts on
the celestial sphere as $SO(3,1)$. It is of interest to perform mode
expansions of free fields in Minkowski spacetime that realize this
symmetry in a simple way. In the present work we perform such a mode
expansion for massive scalar fields using the unitary principal series
representations of $SO(3,1)$ with a view to developing a holographic
approach to gravity in asymptotically flat spacetime. These mode expansions
are also of use in studying holography in three-dimensional de Sitter
spacetime. 
\end{abstract}
\maketitle

\section{Introduction}

There has been considerable interest recently in constructing holographic
theories between flat 4d Minkowski spacetime and a 2d boundary celestial
sphere conformal field theory \citep{deBoer:2003vf,Kapec:2014opa,Kapec:2016jld,Cheung:2016iub}.
Central to this mission is the construction of conformally covariant
wavefunctions that form unitary representations of the ${\rm SO}(d,1)$
group. These wavefunctions are defined on a $d$-dimensional de Sitter
spacetime ${\rm dS}_{d}$, on which ${\rm SO}(d,1)$ acts naturally
through an embedding of ${\rm dS}_{d}$ as a submanifold of a $(d+1)$-dimensional
Minkowski spacetime.

As part of the program to realize the dS/CFT correspondence \citep{Strominger:2001pn},
numerous papers have previously constructed unitary principal series
representations of the ${\rm SO(2,1)}$ group on two-dimensional de
Sitter spacetimes \citep{Guijosa:2003ze,Guijosa:2005qi}, as well
as $q$-deformed versions of the principal series on the three-dimensional
de Sitter spacetime \citep{Lowe:2004nw}. In this paper we construct
the unitary principal series representation of the ${\rm SO(3,1)}$
group on the three-dimensional de Sitter spacetime. We also compute
the uplifted version of these wavefunctions on the ambient four-dimensional
Minkowski spacetime. Finally, we comment on relevant previous work
\citep{Pasterski:2016qvg,Pasterski:2017kqt,Pasterski:2017ylz} and
discuss how our results fit within the program to develop holographic
approaches to gravity in de Sitter and asymptotically Minkowski spacetime.

The sections are organized as follows: we first establish our coordinate
systems and fix our notations in section \ref{sec:coord}. We then
construct massive scalar mode functions on both the three-dimensional
de Sitter spacetime ${\rm dS_{3}}$ and the four-dimensional Minkowski
spacetime ${\rm M_{4}}$, in section \ref{sec:dsmodes} and \ref{sec:minkmodes},
respectively. We then show in section \ref{sec:ups} that these mode
functions form a unitary principal series representation of the ${\rm SO(3,1)}$
group. We note that previous work \citep{Pasterski:2016qvg,Pasterski:2017kqt,Pasterski:2017ylz}
uses modes that form non-unitary highest weight representations of
$SO(3,1)$. Finally we comment in section \ref{sec:further} on how
our mode functions can serve as a conformal basis to develop holographic
formulations of gravity in asymptotically de Sitter and Minkowski
spacetimes.

\section{Conformal Coordinates}

\label{sec:coord}

\begin{figure}
\centering \begin{overpic}[width=0.7\textwidth]{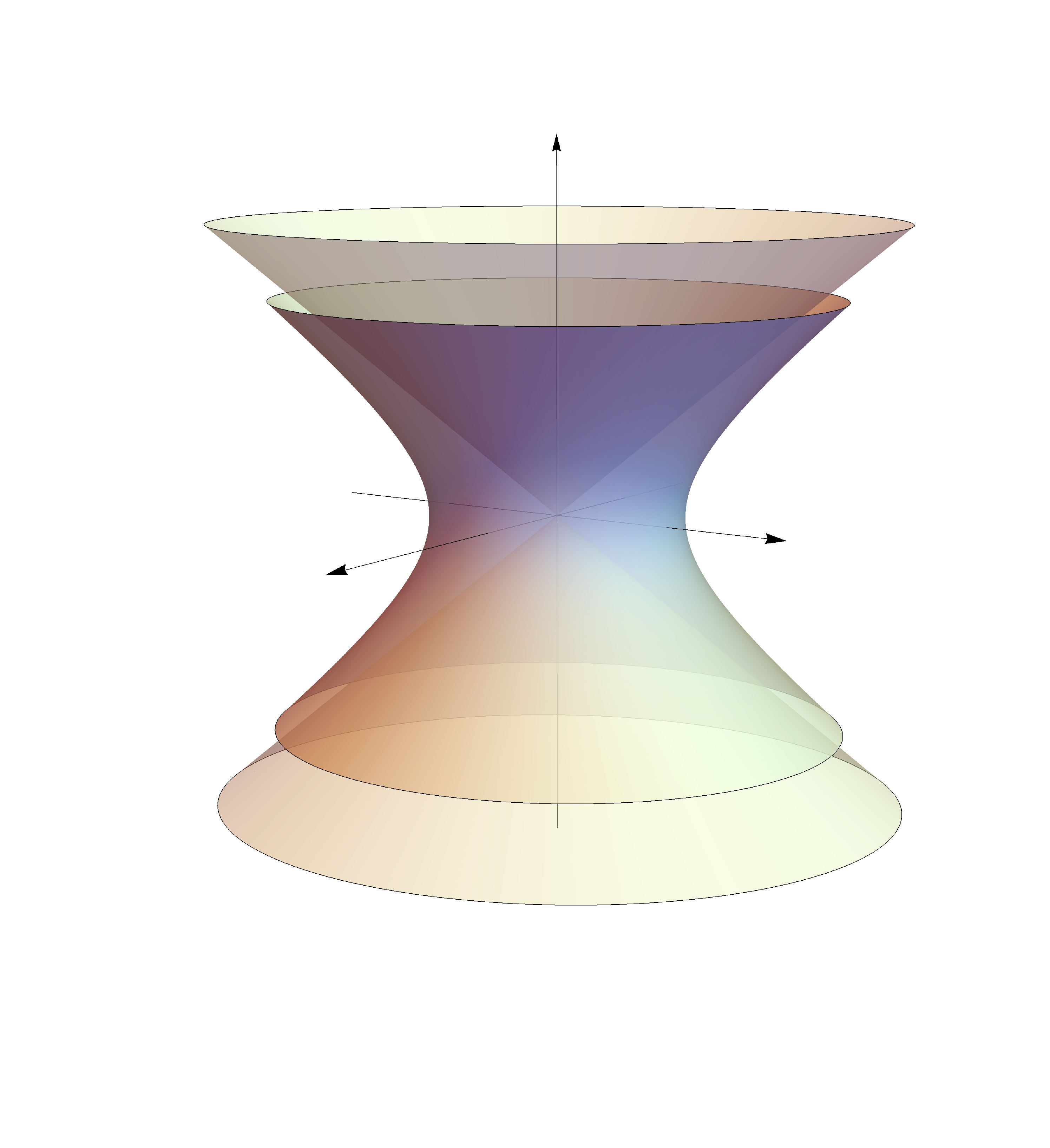} \put
(25,49) {$x^{1}$} \put (69,52) {$\,x^{2},x^{3}$} \put (48,89)
{$x^{0}$} \put (63,59) {$\leftarrow\rho=1$ hypersurface} \put
(17,76) {$\nearrow$} \put (9,73) {$\rho=0$} \put (4,69) {hypersurface}
\end{overpic}

\begin{tikzpicture}[scale=1.5]
  \shade (0,0) -- (1,1) -- (2,0) -- (1,-1) -- (0,0);
  \draw (0,2) node[above] {$i^+$};
  \draw (0,-2) node[below] {$i^-$};
  \draw (2,0) node[right] {$i^0$};
  \draw (1.2,1) node[above] {$\mathcal{I}^+$};
  \draw (1.2,-1) node[below] {$\mathcal{I}^-$};
  \draw (0.7,0) node[right] {$\leftarrow \rho=1$};
  \draw (0,2) -- (0,-2);
  \draw (0,2) -- (2,0);
  \draw (0,-2) -- (2,0);
  \pgfmathsetmacro{\e}{1.5}   % eccentricity
  \pgfmathsetmacro{\a}{0.7}
  \pgfmathsetmacro{\b}{(\a*sqrt((\e)^2-1)}
  \draw plot[domain=-1:1] ({\a*cosh(\x)},{\b*sinh(\x)});
\end{tikzpicture} \caption{Minkowski spacetime may be divided up into radial ($\rho$) slices
isometric to 3d de Sitter spacetime as shown in the top panel. The
bottom panel shows the Penrose diagram of Minkowski spacetime. The
shaded region is the region bounded by $\rho=0$ and $\rho=\infty$.
In particular, $i^{\pm}$ are excluded from this region.}
\label{fig:coord} 
\end{figure}

Before we begin our discussion of the representation theory of ${\rm SO(3,1)}$
in de Sitter and Minkowski spacetimes, we would like to present the
coordinate systems we employ and fix notation. A schematic plot of
the relevant hypersurfaces can be found in Fig.~\ref{fig:coord}.

We start with the 4d flat Minkowski spacetime labeled by coordinates
$(x^{0},x^{1},x^{2},x^{3})$ with the following metric 
\[
\dd s^{2}=-(\dd x^{0})^{2}+(\dd x^{1})^{2}+(\dd x^{2})^{2}+(\dd x^{3})^{2}\,.
\]
We embed our 3d de Sitter spacetime as a hypersurface within the 4d
Minkowski spacetime. To do so we first switch to hyperbolic coordinates
$(t,\rho,\theta,\varphi)$ with $-\infty<t<\infty$, $\rho>0$, $0\le\theta<\pi$
and $0\le\varphi<2\pi$, defined by 
\begin{align}
x^{0} & =\rho\sinh(t/\rho)\nonumber \\
x^{1} & =\rho\cos\theta\cosh(t/\rho)\nonumber \\
x^{2} & =\rho\sin\theta\cos\varphi\cosh(t/\rho)\nonumber \\
x^{3} & =\rho\sin\theta\sin\varphi\cosh(t/\rho)\,.\label{eq:embedding}
\end{align}
Note that these coordinates only cover the region of the Minkowski
spacetime defined by points $x^{\mu}$ where $x\cdot x>0$. The metric
then becomes 
\[
\dd s^{2}=-\dd t^{2}+\frac{2t\,\dd\rho\,\dd t}{\rho}+\left(1-\frac{t^{2}}{\rho^{2}}\right)\dd\rho^{2}+\rho^{2}\cosh^{2}\left(\frac{t}{\rho}\right)\left(\dd\theta^{2}+\sin^{2}\theta\,\dd\varphi^{2}\right)\,.
\]
The d'Alembertian in this coordinate system looks rather complicated.
We therefore perform the following change of variables 
\[
\eta=\frac{t}{\rho}
\]
and the metric takes the following simpler form 
\[
\dd s^{2}=-\rho^{2}\dd\eta^{2}+\dd\rho^{2}+\rho^{2}\cosh^{2}\eta\left(\dd\theta^{2}+\sin^{2}\theta\,\dd\varphi^{2}\right)\,.
\]

A 3d de Sitter spacetime can then be embedded into this 4d Minkowski
spacetime as a hypersurface of constant $\rho=\ell$, where $\ell$
is the de Sitter length scale which we will set to unity in what follows.
On the 3d de Sitter spacetime the induced metric is simply 
\[
\dd s^{2}=-\dd t^{2}+\cosh^{2}t\left(\dd\theta^{2}+\sin^{2}\theta\,\dd\varphi^{2}\right)\,.
\]
Through a stereographic projection we can parameterize the 2-sphere
covered by coordinates $(\theta,\varphi)$ with a complex variable
$z$, and obtain the Fubini-Study metric on the 2-sphere: 
\[
\dd\theta^{2}+\sin^{2}\theta\,\dd\varphi^{2}=\frac{4\,\dd z\,\dd\bar{z}}{(1+|z|^{2})^{2}}\,.
\]
This allows us to rewrite the 3d de Sitter metric as
\begin{equation}
\dd s^{2}=-\dd t^{2}+\cosh^{2}t\,\frac{4\,\dd z\,\dd\bar{z}}{(1+|z|^{2})^{2}}\label{eq:3dds}
\end{equation}
and the 4d Minkowski metric in hyperbolic coordinates as 
\[
\dd s^{2}=-\rho^{2}\dd\eta^{2}+\dd\rho^{2}+\rho^{2}\cosh^{2}\eta\,\frac{4\,\dd z\,\dd\bar{z}}{(1+|z|^{2})^{2}}\,.
\]

\section{3d de Sitter Mode Functions}

\label{sec:dsmodes}

We start from the metric on the 3d de Sitter spacetime \eqref{eq:3dds}.
The isometry group of the 3d de Sitter spacetime is ${\rm SO(3,1)}$
which has 6 generators with real coefficients. The first step in building
a unitary representation of ${\rm SO(3,1)}$ on 3d de Sitter is to
solve the scalar field equation with mass $\mu$ 
\[
(\Delta-\mu^{2})\phi(t,z,\bar{z})=0
\]
where the d'Alembertian is defined in general as 
\[
\Delta\phi=\frac{1}{\sqrt{|g|}}\partial_{i}(g^{ij}\sqrt{|g|}\partial_{j}\phi)\,.
\]
Here and in what follows we use Latin indices when we are referring
to the 3d de Sitter submanifold and use Greek indices when we are
dealing with the ambient 4d Minkowski spacetime. The d'Alembertian
is computed to be 
\[
\Delta\phi=\left[-\partial_{t}^{2}-2(\tanh t)\partial_{t}+(\sech^{2}t)(1+|z|^{2})^{2}\partial_{z}\partial_{\bar{z}}\right]\phi\,.
\]
The massive scalar field equation therefore is 
\[
\left(-\frac{\partial^{2}}{\partial t^{2}}-2\tanh t\frac{\partial}{\partial t}+\mathrm{sech}^{2}t\,(1+|z|^{2})^{2}\frac{\partial^{2}}{\partial z\partial\bar{z}}-\mu^{2}\right)\phi(t,z,\bar{z})=0\,.
\]
We can perform separation of variables as usual and write a mode function
as 
\[
\phi_{lm}(t,z,\bar{z})=\phi_{l}(t)\,Y_{m}^{l}(z,\bar{z})
\]
where $m$ is a $j_{3}$ eigenvalue. The $Y_{m}^{l}$ are the standard
spherical harmonics \citep{kelvin1867treatise} in $(z,\bar{z})$
coordinates. They satisfy the eigenvalue equation 
\[
(1+|z|^{2})^{2}\frac{\partial^{2}}{\partial z\partial\bar{z}}Y_{m}^{l}=-l(l+1)Y_{m}^{l}\,.
\]
Explicitly, we have 
\[
Y_{m}^{l}(\theta,\varphi)=\sqrt{\frac{(2l+1)(l-m)!}{4\pi(l+m)!}}P_{l}^{m}(\cos\theta)e^{im\varphi}
\]
where 
\[
P_{l}^{m}(x)=\frac{(1+x)^{m/2}}{(1-x)^{m/2}}\sum_{k=0}^{l}\frac{\left(\frac{1-x}{2}\right)^{k}(-l)_{k}(l+1)_{k}}{k!\,\Gamma(k-m+1)}\,.
\]
The coordinate transformation linking $(\theta,\varphi)$ and $(z,\bar{z})$
is 
\[
(\sin\theta\,e^{i\varphi},\cos\theta)=\left(\frac{2z}{1+|z|^{2}},\frac{1-|z|^{2}}{1+|z|^{2}}\right)\,.
\]
It is easy to see that the spherical harmonics $Y_{l}^{m}(z,\bar{z})$
can be built out of homogeneous polynomials of the following three
variables 
\[
F_{z}=\frac{2z}{1+|z|^{2}}\qquad F_{\bar{z}}=\frac{2\bar{z}}{1+|z|^{2}}\qquad F_{t}=\frac{1-|z|^{2}}{1+|z|^{2}}\,.
\]
We will later see that these three variables will appear in (\ref{eq:cartesian}).
One is then left to solve 
\[
\left(\frac{\partial^{2}}{\partial t^{2}}+2\tanh t\frac{\partial}{\partial t}+l(l+1)\mathrm{sech}^{2}t+\mu^{2}\right)\phi_{l}(t)=0\,.
\]
This has the following two linearly independent solutions 
\begin{equation}
\phi_{l,1}(t)=\mathrm{sech}t\,P_{l}^{\sqrt{1-\mu^{2}}}\left(\tanh t\right),\quad\phi_{l,2}(t)=\mathrm{sech}t\,Q_{l}^{\sqrt{1-\mu^{2}}}\left(\tanh t\right)\label{eq:basis}
\end{equation}
where $P$ and $Q$ are the associated Legendre functions of first
and second kind, respectively. We expect to get a unitary representation
corresponding to a general complex linear combination, one of which
will be the Euclidean vacuum (see Section \ref{sec:norm} and Appendix
\ref{appendix1}). Other combinations will generate modes around an
$\alpha$-vacuum \citep{Allen:1985ux,Goldstein:2003ut,Goldstein:2003qf}.

\section{Uplifting onto 4d Minkowski}

\label{sec:minkmodes}

To uplift the 3d de Sitter mode functions onto the 4d Minkowski spacetime
we consider the scalar field equation in 4d with mass $M$
\[
(\Box-M^{2})\Phi(\eta,\rho,z,\bar{z})=0\,.
\]
The d'Alembertian is computed to be 
\[
\Box=-\frac{1}{\rho^{2}}(\partial_{\eta}^{2}+2\tanh\eta\,\partial_{\eta})+3\frac{\partial_{\rho}}{\rho}+\partial_{\rho}^{2}+\frac{\sech^{2}\eta\,(1+|z|^{2})^{2}\,\partial_{z}\partial_{\bar{z}}}{\rho^{2}}\,.
\]
Separating variables, the mode functions can be written as
\[
\Phi_{plm}(\eta,\rho,z,\bar{z})=\phi_{pl}(\eta)\,\psi_{p}(\rho)\,Y_{m}^{l}(z,\bar{z})
\]
where $l=0,1,\cdots$, $m=-l,\cdots,l$ and the range of the real
parameter $p$ will be discussed later. We find the differential equation
for $\phi_{pl}$ to be identical to the 3d de Sitter modes, except
that we replace $\mu$ with a real parameter $p$, 
\[
\left(\partial_{\eta}^{2}+2(\tanh\eta)\partial_{\eta}+l(l+1)\sech^{2}\eta+p^{2}\right)\phi_{pl}(\eta)=0
\]
which has solutions \eqref{eq:basis} with the replacement $t\to\eta$
and $\mu\to p$.

The differential equation for $\psi_{p}$ therefore becomes 
\[
\left(\partial_{\rho}^{2}+\frac{3}{\rho}\partial_{\rho}+\frac{p^{2}}{\rho^{2}}-M^{2}\right)\psi_{p}(\rho)=0\,.
\]
This has two independent solutions 
\begin{equation}
\psi_{p,1}=\frac{I_{\sqrt{1-p^{2}}}(M\rho)}{\rho}\,,\qquad\psi_{p,2}=\frac{K_{\sqrt{1-p^{2}}}(M\rho)}{\rho}\label{eq:radialsol}
\end{equation}
where $I_{\alpha}$ and $K_{\alpha}$ are modified Bessel functions
of first and second kind, respectively.

\section{Klein-Gordon Norm and Orthonormality Conditions}

\label{sec:norm}

Before presenting the explicit form of our unitary principal series
representation we would like to first establish the orthonormalizability
of the 4d Minkowski mode functions. The mode functions are normalized
with respect to the Klein-Gordon norm, the most general form of which
is \citep{birrell1984quantum}
\begin{equation}
\langle f,g\rangle=-i\int_{\Sigma}\dd\Sigma\,n^{\lambda}(f\partial_{\lambda}g^{\star}-g^{\star}\partial_{\lambda}f)\,.\label{eq:kgnorm}
\end{equation}
Here $\Sigma$ is a spacelike surface, $n^{\lambda}$ is a timelike
unit vector field normal to $\Sigma$ and $d\Sigma$ is the volume
element in $\Sigma$. This norm is time-independent, and in principle
can be evaluated on any Cauchy slice. Here, for convenience we choose
the $\eta=\eta_{0}$ slice, where $\eta_{0}$ is an arbitrary constant.
On the 4d Minkowski spacetime with coordinate system $(\eta,\rho,\theta,\varphi)$
this then becomes 
\[
\langle f,g\rangle=-i\int_{\rho}\int_{S^{2}}(f\partial_{\eta}g^{\star}-g^{\star}\partial_{\eta}f)|_{\eta=\eta_{0}}\,(\cosh^{2}\eta_{0})\,\rho\,\dd\rho\,\dd\Omega
\]
where $\dd\Omega$ is the area element of the unit 2-sphere $S^{2}$.
Given that our mode functions are separable $\Phi_{plm}(\eta,\rho,z,\bar{z})=\phi_{pl}(\eta)\psi_{p}(\rho)Y_{m}^{l}(z,\bar{z})$
we can further evaluate this norm to obtain 
\begin{align*}
\langle\Phi_{plm},\Phi_{p'l'm'}\rangle= & -i\,\left(\phi_{pl}(\eta_{0})\dot{\phi}_{p'l'}^{\star}(\eta_{0})-\phi_{p'l'}^{\star}(\eta_{0})\dot{\phi}{}_{pl}(\eta_{0})\right)\,\left(\cosh^{2}\eta_{0}\right)\\
 & \qquad\times\int_{0}^{\infty}\psi_{p}(\rho)\psi_{p'}^{\star}(\rho)\,\rho\,\dd\rho\int_{S^{2}}Y_{m}^{l}(z,\bar{z})Y_{m'}^{\star l'}(z,\bar{z})\,\dd\Omega
\end{align*}
where $\dot{\phi}(\eta)=\partial_{\eta}\phi$. The orthonormality
of the spherical harmonics allows us to conclude that 
\[
\int_{S^{2}}Y_{m}^{l}(z,\bar{z})Y_{m'}^{\star l'}(z,\bar{z})\,\dd\Omega=\delta_{ll'}\delta_{mm'}\,.
\]
For the $\rho$ integral, note that multiplying the differential equation
satisfied by $\psi_{p}$ with $\psi_{p'}^{\star}$ gives 
\[
\psi''_{p}\psi_{p'}^{\star}+\frac{3}{\rho}\psi'_{p}\psi_{p'}^{\star}+\left(\frac{p^{2}}{\rho^{2}}-M^{2}\right)\psi_{p}\psi_{p'}^{\star}=0
\]
where $\psi'=\partial_{\rho}\psi$. Likewise swapping $\psi_{p}\leftrightarrow\psi_{p'}^{*}$
and subtracting we have 
\[
-\frac{p^{2}-p'^{2}}{\rho^{2}}\psi_{p}\psi_{p'}^{\star}=\psi''_{p}\psi_{p'}^{\star}-\psi_{p'}^{\star''}\psi_{p}+\frac{3}{\rho}\psi'_{p}\psi_{p'}^{\star}-\frac{3}{\rho}\psi_{p'}^{\star'}\psi_{p}\,.
\]
Integrating, we have 
\[
-(p^{2}-p'^{2})\int_{0}^{\infty}\psi_{p}\psi_{p'}^{\star}\,\rho\,\dd\rho=\int_{0}^{\infty}\dd\rho\,[\rho^{3}(\psi''_{p}\psi_{p'}^{\star}-\psi_{p'}^{\star''}\psi_{p})+3\rho^{2}(\psi'_{p}\psi_{p'}^{\star}-\psi_{p'}^{\star'}\psi_{p})]\,.
\]
The integral on the right hand side can be integrated by parts to
yield 
\begin{equation}
-(p^{2}-p'^{2})\int_{0}^{\infty}\psi_{p}\psi_{p'}^{\star}\,\rho\,\dd\rho=[\rho^{3}(\psi'_{p}\psi_{p'}^{\star}-\psi_{p'}^{\star'}\psi_{p})]\Big|_{0}^{\infty}\,.\label{eq:norm-int}
\end{equation}

The modified Bessel functions have the following mirror symmetry 
\[
I_{\alpha}^{\star}(z)=I_{\alpha^{\star}}(z^{\star})\qquad K_{\alpha}^{\star}(z)=K_{\alpha^{\star}}(z^{\star})\,.
\]
At $z=+\infty$, $I_{\alpha}(z)\sim e^{z}$ which increases exponentially
while $K_{\alpha}(z)\sim e^{-z}$ which decreases exponentially. We
therefore discard the $\psi_{p,1}$ set of solutions as these modes
are not normalizable and study the \eqref{eq:radialsol} solutions
$\psi_{p,2}$. We begin by considering the case $p^{2}>1$ and take
the branch $\sqrt{1-p^{2}}=i\sqrt{p^{2}-1}$. Let us define $\alpha=\sqrt{p^{2}-1}$
and henceforth we will drop the 2 subscript on $\psi_{p,2}$. Near
$z=0$, the expansion 
\[
K_{\nu}(z)=2^{\nu-1}\Gamma(\nu)z^{-\nu}+2^{-\nu-1}\Gamma(-\nu)z^{\nu}+\cdots
\]
allows us to evaluate the surface term to give 
\begin{align}
 & \lim_{\rho\to0}\rho^{3}\left(\psi'_{p}\psi_{p'}^{*}-\psi_{p'}^{*'}\psi_{p}\right)=\nonumber \\
 & \quad\lim_{\rho\to0}i(\alpha+\alpha')2^{-2}\left(\Gamma(i\alpha')\Gamma(-i\alpha)(M\rho/2)^{i(\alpha-\alpha')}-\Gamma(i\alpha)\Gamma(-i\alpha')(M\rho/2)^{-i(\alpha-\alpha')}\right)\label{eq:surfaceterm}
\end{align}
treating the rapidly oscillating terms in $\alpha$ as vanishing in
the sense of a distribution. For $\rho\to0$, this is proportional
to a sinc representation of the Dirac delta function
\[
\lim_{\rho\to0}\rho^{3}\left(\psi'_{p}\psi_{p'}^{*}-\psi_{p'}^{*'}\psi_{p}\right)=-\frac{\alpha\Gamma(i\alpha)\Gamma(-i\alpha)}{2}\sin\left[(\alpha-\alpha')\log(M\rho/2)\right]
\]
which gives 
\[
\int_{0}^{\infty}\psi_{p}\psi_{p'}^{\star}\,\rho\,\dd\rho=\frac{\Gamma(i\alpha)\Gamma(-i\alpha)}{4}\lim_{C\to\infty}\frac{\sin[C(\alpha-\alpha')]}{\alpha-\alpha'}
\]
where we have set $C=\log(M\rho/2)$. Using the following identity
\[
\lim_{C\to\infty}\frac{\sin(Cx)}{x}=\pi\delta(x)
\]
we then have 
\begin{equation}
\int_{0}^{\infty}\psi_{p}\psi_{p'}^{\star}\,\rho\,\dd\rho=\frac{\pi\Gamma(i\alpha)\Gamma(-i\alpha)}{4}\delta(\alpha-\alpha')=\frac{\pi^{2}}{4p\sinh\left(\pi\sqrt{p^{2}-1}\right)}\delta\left(p-p'\right)\,.\label{eq:radialnorm}
\end{equation}
 For $0\le p^{2}<1$, instead of oscillating terms in eq.~\ref{eq:surfaceterm}
we have power-law divergencies, and the mode functions $\psi_{p}$
in this case again are not normalizable.

To conclude, the normalizable radial mode functions arise from $\psi_{p,2}$
in \eqref{eq:radialsol} with $p^{2}>1$. The $p$-dependent prefactor
in \eqref{eq:radialnorm} may then be absorbed into the normalization
of these functions. We assume this has been done, and in the interest
of notational clarity we will from now use $\psi_{p}$ to denote the
normalized radial mode functions. The normalized radial mode functions
in this case will satisfy the following 
\begin{equation}
\int_{0}^{\infty}\psi_{p}\psi_{p'}^{\star}\,\rho\,\dd\rho=\delta(p-p')\,.\label{eq:radialinnerprod}
\end{equation}

For the $\eta$ dependence, we form the following linear combination
of $\phi_{pl,1}$ and $\phi_{pl,2}$ to obtain 
\[
\phi_{pl}=\frac{i\pi}{2}\phi_{pl,1}+\phi_{pl,2}
\]
which, as we will show in Appendix \ref{appendix1}, are the (unnormalized)
modes corresponding to the 4d Minkowski vacuum. Using 
\[
\frac{\partial P_{\nu}^{\mu}}{\partial z}=\frac{\nu zP_{\nu}^{\mu}(z)-(\mu+\nu)P_{\nu-1}^{\mu}(z)}{z^{2}-1}\qquad\frac{\partial Q_{\nu}^{\mu}}{\partial z}=\frac{\nu zQ_{\nu}^{\mu}(z)-(\mu+\nu)Q_{\nu-1}^{\mu}(z)}{z^{2}-1}
\]
and 
\[
P_{\nu}^{\mu}(0)=\frac{\pi^{1/2}2^{\mu}}{\Gamma\left(\frac{1-\mu-\nu}{2}\right)\Gamma\left(\frac{\nu-\mu}{2}+1\right)}\qquad Q_{\nu}^{\mu}(0)=-\frac{\pi^{3/2}2^{\mu-1}\tan\frac{\pi(\mu+\nu)}{2}}{\Gamma\left(\frac{1-\mu-\nu}{2}\right)\Gamma\left(\frac{\nu-\mu}{2}+1\right)}
\]
we can evaluate the $\eta$-dependent part of the Klein-Gordon norm
to obtain 
\[
-i[\phi_{pl}(\eta_{0})\phi_{pl}^{\star'}(\eta_{0})-\phi_{pl}^{\star}(\eta_{0})\phi'_{pl}(\eta_{0})](\cosh^{2}\eta_{0})=\pi e^{-\pi\sqrt{p^{2}-1}}
\]
which does not depend on $\eta_{0}$ due to conservation of the Klein-Gordon
norm. We can therefore normalize the modes $\phi_{pl}$ by replacing
\[
\phi_{pl}\to\frac{e^{\frac{\pi}{2}\sqrt{p^{2}-1}}}{\pi^{1/2}}\phi_{pl}\,.
\]
In what follows we will assume that this has been done and in the
interest of notational simplicity we will use $\phi_{pl}$ to denote
the normalized modes.

To summarize, for $p^{2}>1$ we have constructed mode functions of
the 4d Klein-Gordon equation corresponding to the Minkowski vacuum,
which when restricted to the de Sitter slice $\rho=1$ correspond
to the Euclidean vacuum of the 3d de Sitter spacetime. These modes
$\Phi_{plm}(\eta,\rho,z,\bar{z})=\phi_{pl}(\eta)\psi_{p}(\rho)Y_{m}^{l}(z,\bar{z})$
are normalized with respect to to the Klein-Gordon norm (\ref{eq:kgnorm})
with the following orthonormality condition 
\[
\langle\Phi_{plm},\Phi_{p'l'm'}\rangle=\delta(p-p')\delta_{ll'}\delta_{mm'}\,.
\]
For $0\le p^{2}<1$, the radial mode functions are not normalizable.

\section{Unitary Principal Series Representation}

\label{sec:ups} We are now in a position to build the unitary principal
series representation of ${\rm SO(3,1)}$ which acts on the ${\rm dS_{3}}$/${\rm M_{4}}$
mode functions. Since the action of ${\rm SO(3,1)}$ on ${\rm M_{4}}$
leaves the radial coordinate $\rho$ invariant, the actions is identical
on both ${\rm dS_{3}}$ mode functions and on ${\rm M_{4}}$ mode
functions. For simplicity of presentation we will focus on ${\rm dS_{3}}$
modes, but all equations in this section carry over to the ${\rm M_{4}}$
modes trivially.

On the past and future null infinities $\mathcal{I^{\pm}}$ of the
3d de Sitter spacetime, the Killing vectors can be written as conformal
Killing vectors of the spatial 2-sphere 
\[
L_{n}=-z^{n+1}\frac{\partial}{\partial z},\,\bar{L}_{n}=-\bar{z}^{n+1}\frac{\partial}{\partial\bar{z}}
\]
where $n=0,\pm1$. However one should use caution in applying this
formula. It is correct when acting on the metric, or massless scalars,
but as we will see there will be additional terms that must be added
depending on the class of functions or fields considered. General
complex combinations of these vectors will not preserve the desired
reality conditions, so we will need to be careful to construct the
correct 6 independent generators that will appear with real coefficients.

To extend these into the bulk of de Sitter it is helpful to arrange
them into an ${\rm SO(3)}$ corresponding to the isometries of the
spatial slices. With the convention $g=\exp(i\theta J)$ we find the
generators: 
\begin{align*}
J_{1} & =\frac{i}{2}\left(L_{-1}+L_{1}+\bar{L}_{-1}+\bar{L}_{1}\right)\\
J_{2} & =\frac{1}{2}\left(L_{-1}-L_{1}-\bar{L}_{-1}+\bar{L}_{1}\right)\\
J_{3} & =L_{0}-\bar{L}_{0}\,.
\end{align*}
These immediately extend into the bulk without time dependent contributions.
This also allows us to read off the conjugation condition to be imposed
on the generators. Since the $J_{k}$ are Hermitian we require 
\[
L_{n}^{\dagger}=-\bar{L}_{n},\qquad\bar{L}_{n}^{\dagger}=-L_{n}\,.
\]

The time-dependent Killing vectors take the form 
\[
K=F\frac{\partial}{\partial t}+\frac{1}{2}\left(1+z\bar{z}\right)^{2}\tanh t\,\left(\partial_{\bar{z}}F\right)\frac{\partial}{\partial z}+\frac{1}{2}\left(1+z\bar{z}\right)^{2}\tanh t\,\left(\partial_{z}F\right)\frac{\partial}{\partial\bar{z}}
\]
where $F$ is one of the three solutions 
\begin{equation}
F_{z}=\frac{2z}{1+|z|^{2}}\qquad F_{\bar{z}}=\frac{2\bar{z}}{1+|z|^{2}}\qquad F_{t}=\frac{1-|z|^{2}}{1+|z|^{2}}\,.\label{eq:cartesian}
\end{equation}
At $\mathcal{I}^{+}$ these reduce to 
\begin{align*}
\tilde{K}_{1} & =L_{1}-\bar{L}_{-1}\\
\tilde{K}_{2} & =\bar{L}_{1}-L_{-1}\\
\tilde{K}_{3} & =L_{0}+\bar{L}_{0}
\end{align*}
when acting on the metric. It is convenient to assemble these into
Hermitian linear combinations: 
\begin{align*}
K_{1} & =\frac{1}{2}\left(L_{1}-\bar{L}_{-1}-\bar{L}_{1}+L_{-1}\right)\\
K_{2} & =\frac{i}{2}\left(\bar{L}_{1}-L_{-1}+L_{1}-\bar{L}_{-1}\right)\\
K_{3} & =-i\left(L_{0}+\bar{L}_{0}\right)
\end{align*}
which at a general spacetime point become 
\begin{align}
K_{1} & =\frac{1}{2}\left(\frac{2(z-\bar{z})}{1+|z|^{2}}\frac{\partial}{\partial t}-\tanh t\left(\left(z^{2}+1\right)\frac{\partial}{\partial z}-\left(\bar{z}^{2}+1\right)\frac{\partial}{\partial\bar{z}}\right)\right)\nonumber \\
K_{2} & =\frac{i}{2}\left(\frac{2(z+\bar{z})}{1+|z|^{2}}\frac{\partial}{\partial t}-\tanh t\left(\left(z^{2}-1\right)\frac{\partial}{\partial z}+\left(\bar{z}^{2}-1\right)\frac{\partial}{\partial\bar{z}}\right)\right)\nonumber \\
K_{3} & =-i\left(\frac{1-|z|^{2}}{1+|z|^{2}}\frac{\partial}{\partial t}-\tanh t\left(z\frac{\partial}{\partial z}+\bar{z}\frac{\partial}{\partial\bar{z}}\right)\right)\,.\label{eq:bulkgen}
\end{align}
Note the generators satisfy the canonical Lorentz algebra 
\begin{align*}
[J_{i},J_{j}] & =i\epsilon_{ijk}J_{k}\\{}
[J_{i},K_{j}] & =i\epsilon_{ijk}K_{k}\\{}
[K_{i},K_{j}] & =-i\epsilon_{ijk}J_{k}\,.
\end{align*}

On the three-dimensional de Sitter mode functions (\ref{eq:basis})
the generators (\ref{eq:bulkgen}) take the simplified form 
\begin{align}
K_{1} & =\frac{1}{2}\left(\frac{2(z-\bar{z})}{1+|z|^{2}}\left(2h_{+}-2\right)-\left(\left(z^{2}+1\right)\frac{\partial}{\partial z}-\left(\bar{z}^{2}+1\right)\frac{\partial}{\partial\bar{z}}\right)\right)\nonumber \\
K_{2} & =\frac{i}{2}\left(\frac{2(z+\bar{z})}{1+|z|^{2}}\left(2h_{+}-2\right)-\left(\left(z^{2}-1\right)\frac{\partial}{\partial z}+\left(\bar{z}^{2}-1\right)\frac{\partial}{\partial\bar{z}}\right)\right)\nonumber \\
K_{3} & =-i\left(\frac{1-|z|^{2}}{1+|z|^{2}}\left(2h_{+}-2\right)-\left(z\frac{\partial}{\partial z}+\bar{z}\frac{\partial}{\partial\bar{z}}\right)\right)\label{eq:bulkgenprin}
\end{align}
where we define 
\[
2h_{\pm}=\frac{d-1}{2}\pm\sqrt{\left(\frac{d-1}{2}\right)^{2}-\mu^{2}}=1\pm\sqrt{1-\mu^{2}}\,.
\]
To check this we note that the group of rotations ${\rm SO(3)}$ acts
straightforwardly in the basis \eqref{eq:basis}. The rotation generators
also rotate the $K_{i}$ amongst themselves, so we can focus on the
action of say $K_{3}$ on the mode functions \eqref{eq:basis}. It
is then straightforward to check that 
\[
\frac{\partial}{\partial t}\phi_{l=0}=2(h_{+}-1)\phi_{l=1}
\]
which holds for the solutions $\phi_{l,1}$ and $\phi_{l,2}$ of \eqref{eq:basis}
independently and determines the prefactors that appear in \eqref{eq:bulkgenprin}.
Likewise, when acting on the four-dimensional Minkowski modes the
generators take the exact same expression with $\mu$ replaced by
$p$.

To confirm the generators match the principal series we will use the
representation of ${\rm SO(3,1)}$ on functions $L^{2}(S^{2})$. Note
this differs from the more common representation on functions $L^{2}(\mathbb{C})$,
which would be applicable to the flat slicing of de Sitter. Likewise
there is a representation on $L^{2}(H^{2})$, though we won't need
that here. The upshot of these different realizations of the principal
series is that the extra terms in \eqref{eq:bulkgenprin} take completely
different forms.

To realize the representation we consider the cone $C_{+}^{3}$ embedded
in 4d Minkowski spacetime, where 
\[
x_{1}^{2}+x_{2}^{2}+x_{3}^{2}-x_{0}^{2}=0\,.
\]
We then consider the slice through the cone where $x_{0}=1$. This
slice is an $S^{2}$ which may be parameterized by coordinates $z$
above using the Fubini-Study metric. The cone maps into itself under
${\rm SO(3,1)}$. The principal series may be defined as functions
on the slice that behave as \citep{vilenkin1992representation} 
\begin{equation}
(T^{\sigma}(g)f)(z)=\alpha(z,g)^{\sigma}f\left(\frac{g^{-1}\cdot z}{\alpha(z,g)}\right)\label{eq:prindef}
\end{equation}
where $g$ is a ${\rm SO(3,1)}$ group element, and $\alpha(z,g)$
is defined to be the rescaling factor needed to return $g^{-1}\cdot x^{\mu}$
to the slice $x_{0}=1$. The action of ${\rm SO(3,1)}$ on $z$ is
the usual fraction linear transformation, but the factor $\alpha$
depends on which realization of the principal series we are considering.
To write the action of ${\rm SL}(2,\mathbb{C})$ on the Minkowski
coordinates it is helpful to use the familiar representation 
\[
x^{\mu}=\frac{1}{2}\mathrm{Tr}\left(M\sigma^{\mu}\right),\qquad M=\left(\begin{array}{cc}
x^{0}+x^{3} & x^{1}-ix^{2}\\
x^{1}+ix^{2} & x^{0}-x^{3}
\end{array}\right),\qquad\sigma^{\mu}=\left(\mathbbm{1},\sigma^{i}\right)
\]
where $\sigma^{i}$ are the Pauli matrices. Then an ${\rm SL}(2,\mathbb{C})$
transformation acts as 
\[
M'=SMS^{\dagger}\,.
\]
To rescale back to the slice $x^{0}=1$ we rescale to 
\[
\tilde{M}'=\frac{M'}{\frac{1}{2}\mathrm{Tr}\left(M'\right)}\,.
\]
Finally, the coordinates on the 2-sphere $x^{0}=1$ are matched with
the Fubini-Study coordinates via 
\begin{align*}
x^{1}+ix^{2} & =\frac{2z}{1+|z|^{2}}\\
x^{1}-ix^{2} & =\frac{2\bar{z}}{1+|z|^{2}}\\
x^{3} & =\frac{1-|z|^{2}}{1+|z|^{2}}\,.
\end{align*}
From these equations we can read off the factor $\alpha$ and determine
the action of the $z$ coordinate. For 
\[
S=\left(\begin{array}{cc}
a & b\\
c & d
\end{array}\right)\,,
\]
the fractional linear transformation is 
\[
z'=\frac{dz+c}{bz+a}\,.
\]

Since we know the group of rotations acts in a straightforward way
on the spherical harmonics, it suffice to consider one of the boost
generators to check the matching of the generators. To do this we
Taylor expand \eqref{eq:prindef} for a group element of the form
$g=\exp(ik_{3}\epsilon)$. Plugging in the above relations gives 
\[
K_{3}=-i\left(\frac{1-|z|^{2}}{1+|z|^{2}}\sigma-\left(z\frac{\partial}{\partial z}+\bar{z}\frac{\partial}{\partial\bar{z}}\right)\right)\,.
\]
We therefore identify the scalar field representation with a principal
series representation where $\sigma=-2h_{-}$. We have $\sigma=-1+i\sqrt{\mu^{2}-1}$
in the notation of \citep{vilenkin1992representation}. We note the
equivalence of the representations under the replacement $h_{+}\to h_{-}$
and $2h_{+}-2=-2h_{-}$. For the principal series, the inner product
is simply the usual integral over the 2-sphere in Fubini-Study coordinates
which matches the Klein-Gordon norm up to a constant factor. For $0<\mu^{2}<1$
we have the complementary series representations and the above results
extend straightforwardly to that case.

\section{Relation to Previous Work}

\label{sec:psrep}

Here we show the mode functions computed in Ref.~\citep{Pasterski:2017kqt}
form a non-unitary highest-weight representation, and therefore do
not produce a unitary principal series representation of $SO(3,1)$.
To do so we first show that the generators of the special conformal
transformations annihilate a mode function corresponding to the highest
weight of the representation. The mode functions (eq.~2.19 of \citep{Pasterski:2017kqt})
are parameterized by the tuple $(\Delta,\vec{w})$ where $\Delta$
is in general a complex number and $\vec{w}=(w_{x},w_{y})\in\mathbb{R}^{2}$.
These mode functions are 
\begin{equation}
\phi_{\Delta}^{\pm}(X^{\mu};\vec{w})=\frac{4\pi}{(im)}\frac{(\sqrt{-X^{2}})^{\Delta-1}}{(-q(\vec{w})\cdot X\mp i\epsilon)^{\Delta}}K_{\Delta-1}(m\sqrt{X^{2}})\,.\label{eq:psmode}
\end{equation}
Here $X^{\mu}$ are the usual flat coordinates of the 4d Minkowski
spacetime ${\rm M_{4}}$. The $q^{\mu}(\vec{w})$ is the following
4-vector 
\[
q^{\mu}(\vec{w})=(1+|\vec{w}|^{2},2\vec{w},1-|\vec{w}|^{2})\,.
\]
The conformal group ${\rm SO(3,1)}$ acts on the space of scalar functions
defined on ${\rm M_{4}\times\mathbb{R}^{2}}$ by acting on ${\rm M_{4}}$
with the usual Lorentz transformation and on $\mathbb{R}^{2}$ with
the 2D conformal transformations (2D translations, 2D rotations, dilatations
and special conformal transformations): 
\[
\phi_{\Delta}(X^{\mu};\vec{w})\to\phi_{\Delta}(\Lambda^{\mu}{}_{\nu}X^{\nu};\vec{w}'(\vec{w}))\,.
\]
Here $\Lambda^{\mu}{}_{\nu}$ is the Lorentz transformation corresponding
to the ${\rm SO(3,1)}$ group element, and the $\vec{w}'(\vec{w})$
is the conformal transformation corresponding to the ${\rm SO(3,1)}$
element. For special conformal transformation, it is 
\[
\vec{w}'=\frac{\vec{w}+|\vec{w}|^{2}\vec{b}}{1+2\vec{b}\cdot\vec{w}+|\vec{b}|^{2}|\vec{w}|^{2}}\,.
\]
These are labeled by a vector $\vec{b}\in\mathbb{R}^{2}$.

Consider an infinitesimal group element near the identity 
\[
\Lambda^{\mu}{}_{\nu}=\delta_{\nu}^{\mu}+(\delta\Lambda)^{\mu}{}_{\nu}
\]
this has the corresponding infinitesimal transformation on $\mathbb{R}^{2}$
\[
\vec{w}'=\vec{w}+\delta\vec{w}\,.
\]
In particular, for the special conformal transformation parameterized
by an infinitesimal $\vec{b}$ we have the infinitesimal transformation
\[
\vec{w}'=\vec{w}-2(\vec{w}\cdot\vec{b})\vec{w}+|\vec{w}|^{2}\vec{b}\,.
\]
Following the conventions of \citep{Pasterski:2017kqt} this corresponds
to the infinitesimal Lorentz transformation where 
\[
(\delta\omega)^{0}{}_{i}=(\delta\omega)^{i}{}_{0}=(\delta\omega)^{3}{}_{i}=-(\delta\omega)^{i}{}_{3}=b_{i}
\]
with all other components being zero. Here $i=x,y$ labels the indices
of $\mathbb{R}^{2}$. In other words we have 
\[
\begin{bmatrix}X'^{0}\\
X'^{i}\\
X'^{3}
\end{bmatrix}=\begin{bmatrix}X^{0}\\
X^{i}\\
X^{3}
\end{bmatrix}+\begin{bmatrix}b_{i}X^{i}\\
b^{i}(X^{0}-X^{3})\\
b_{i}X^{i}
\end{bmatrix}\,.
\]
Taylor-expanding $\phi_{\Delta}(\Lambda^{\mu}{}_{\nu}X^{\nu};\vec{w}'(\vec{w}))$
and substituting the expressions for $\delta\omega$ and $\delta\vec{w}$
above, we have 
\begin{align*}
\phi_{\Delta}(\Lambda^{\mu}{}_{\nu}X^{\nu};\vec{w}'(\vec{w})) & =\phi_{\Delta}(X^{\mu}+\left(\delta\omega\right)_{\enskip\nu}^{\mu}X^{\nu};\vec{w}+\delta\vec{w})\\
 & =\phi_{\Delta}(X^{\nu};\vec{w})+\left(\delta\omega\right)_{\enskip\nu}^{\mu}X^{\nu}\left(\frac{\partial}{\partial X^{\mu}}\phi_{\Delta}\right)+\delta\vec{w}\cdot\left(\frac{\partial}{\partial\vec{w}}\phi_{\Delta}\right)\\
 & =\phi_{\Delta}(X^{\nu};\vec{w})+b_{i}X^{i}\left(\frac{\partial}{\partial X^{0}}+\frac{\partial}{\partial X^{3}}\right)\phi_{\Delta}+b^{i}(X^{0}-X^{3})\left(\frac{\partial}{\partial X^{i}}\phi_{\Delta}\right)\\
 & \qquad+\left(-2(\vec{w}\cdot\vec{b})\vec{w}+|\vec{w}|^{2}\vec{b}\right)\cdot\left(\frac{\partial}{\partial\vec{w}}\phi_{\Delta}\right)\,.
\end{align*}
For infinitesimal $\vec{b}$ this evaluates to
\begin{equation}
\delta\phi_{\Delta}=\frac{8\pi\Delta}{im}\vec{b}\cdot\vec{w}\frac{\left(\sqrt{-X^{2}}\right){}^{\Delta-1}}{(-q(\vec{w})\cdot X\mp i\epsilon)^{\Delta}}K_{\Delta-1}\left(m\sqrt{X^{2}}\right)\,.\label{eq:lxpsmode}
\end{equation}
In particular, the special conformal transformations annihilate the
mode functions when the weight $\vec{w}=(0,0)$. This implies \citep{Simmons-Duffin:2016gjk}
that the mode functions (\ref{eq:psmode}) form a highest-weight representation
of $SO(3,1)$, and since $SO(3,1)$ has only non-unitary highest-weight
representations, these mode functions cannot form a unitary principal
series.

\section{Discussion}

\label{sec:further} The results of the previous sections lead to
a proposal for a holographic mapping between the 3D de Sitter modes
and conformal operators on a two-sphere. Likewise, the construction
may be lifted to 4D Minkowski spacetime.

\begin{comment}
In the context of celestial sphere amplitudes \citep{deBoer:2003vf,Kapec:2014opa,Kapec:2016jld,Cheung:2016iub}
it is tempting to use our 4D Minkowski spacetime mode functions as
a conformal basis of wavefunctions against which all conformally invariant
amplitudes can be decomposed. 
\end{comment}

\subsection{Holographic mapping between 3D de Sitter spacetime and a Euclidean
2-sphere}

For the case of a flat slicing, it was possible to define a bulk-to-boundary
map and its inverse map \citep{Chatterjee:2015pha} via a construction
reminiscent of the LSZ reduction formula in asymptotically flat spacetime
\citep{Lehmann:1954rq}. That construction does not extend immediately
to the case of the sphere slicing, but we will see the Klein-Gordon
inner product defined above can still be used to extract a natural
set of operators living on the 2-sphere.

We work with a scalar bulk field of mass $\mu$
\begin{equation}
\phi(t,z,\bar{z})=\sum_{l=0}^{\infty}\sum_{m=-l}^{l}a_{lm}\phi_{lm}(t,z,\bar{z})+a_{lm}^{\dagger}\phi_{lm}^{\dagger}(t,z,\bar{z})\label{eq:modeexpan}
\end{equation}
where $\phi_{lm}$ is defined using the Euclidean vacuum modes \eqref{eq:evacmode}
and $a_{lm}$ and $a_{lm}^{\dagger}$ are standard creation/annihilation
operators. Let us consider a late-time sphere at $t=T$ and build
the following inner product using the Klein-Gordon inner product in
3d de Sitter spacetime
\[
\mathcal{O}_{lm}=\left\langle \phi_{lm}(t),\phi(t,z,\bar{z})\right\rangle _{t=T}
\]
which identifies $\mathcal{O}_{lm}=a_{lm}$. Likewise one can define
$\mathcal{O}_{lm}^{\dagger}=a_{lm}^{\dagger}$. One may view this
mapping as a holographic map, with $l,m$ being dual variables to
the coordinates on the 2-sphere $z,\bar{z}$. The formula \eqref{eq:modeexpan}
can then be viewed as the inverse mapping reconstructing the bulk
field in terms of boundary operators. The boundary operators will
obey their usual commutation relations, and at the same time will
transform as a representation of the unitary principal series as described
above. All the above is established at the level of free field theory.
Once interactions are included it seems difficult to view the resulting
boundary theory as any kind of conventional field theory \citep{Chatterjee:2016ifv}.

\subsection{Holographic Mapping Between Celestial Sphere and 4D Minkowski Spacetime}

This procedure can be extended to the 4D Minkowski case. However we
will now have a continuous spectrum of allowed conformal weights $\Delta$
corresponding to the continuous spectrum of radial quantum numbers
$p$. For a scalar field we can use the orthogonality of the radial
mode functions \eqref{eq:radialinnerprod} to project onto a particular
$p$ eigenvalue and then follow the procedure of the previous subsection
to build a boundary operator. The mode expansion of the bulk field
is now
\[
\Phi(\eta,\rho,z,\bar{z})=\sum_{l=0}^{\infty}\sum_{m=-l}^{l}\int_{1}^{\infty}dp\left(a_{plm}\Phi_{plm}(\eta,\rho,z,\bar{z})+a_{plm}^{\dagger}\Phi_{plm}^{\dagger}(\eta,\rho,z,\bar{z})\right)
\]
where $a_{plm}$ and $a_{plm}^{\dagger}$ are annihilation and creation
operators. Again we can define an operator on the celestial 2-sphere
by constructing
\[
\mathcal{O}_{\Delta_{p},lm}=\left\langle \Phi_{plm}(\eta,\rho,z,\bar{z}),\Phi(\eta,\rho,z,\bar{z})\right\rangle =a_{plm}
\]
where the operator transforms as a unitary principal series representation
parameterized by $\Delta_{p}$. Likewise one may define a conjugate
operator $\mathcal{O}_{\Delta_{p},lm}^{\dagger}=a_{plm}^{\dagger}$.
We can therefore interpret this as a holographic map between the bulk
4D Minkowski spacetime and the boundary 2D celestial sphere. One ends
up with a continuous family of boundary operators labelled by the
radial quantum number $p$. As above, it is not clear how the construction
extends to interacting theories.

\appendix
%dummy comment inserted by tex2lyx to ensure that this paragraph is not empty

\section{Euclidean Vacuum}

\label{appendix1} The closed form expression of the positive-frequency
Euclidean modes has been computed in Ref.~\citep{Bousso:2001mw},
eq.~3.37, which we reproduce below in the interest of being self-contained.
Translating the notations of Ref.~\citep{Bousso:2001mw} to our notations,
the time-dependent component of the (unnormalized) Euclidean modes
of the 3d de Sitter spacetime is 
\begin{equation}
\phi_{l}^{E}(t)=(\cosh^{l}t)\,e^{(l+1+i\sqrt{\mu^{2}-1})t}\,{}_{2}{\rm F}_{1}(l+1,l+1+i\sqrt{\mu^{2}-1};2l+2;1+e^{2t})\,.\label{eq:evacmode}
\end{equation}
 The identities \citep{pident} and \citep{qident} may be used provided
we continue $t$ to the complex plane. It is then straightforward
to verify that the linear combination of modes in \eqref{eq:basis}

\[
\phi_{l}^{E}(t)\propto\frac{i\pi}{2}\phi_{l,1}(t)+\phi_{l,2}(t)
\]
are the (unnormalized) positive-frequency modes corresponding to the
Euclidean vacuum of the 3d de Sitter spacetime upon continuing $t$
to real values. These when uplifted to the 4d Minkowski spacetime
will correspond to the Minkowski vacuum, since both are distinguished
by the fact that their Wightman function has a Hadamard singularity
\citep{cmp/1103904566}. The normalization factor is determined in
Section \ref{sec:norm}.

\bibliographystyle{utphys}
\bibliography{ups}

\end{document}